\documentclass[12pt, preprint]{aastex}
\usepackage{amsmath}

\begin{document}
\title{Understanding TeV-band cosmic-ray anisotropy}
\author{Martin Pohl\altaffilmark{1,2},
David Eichler\altaffilmark{3}}
\altaffiltext{1}{DESY, 15738 Zeuthen, Germany; pohlmadq@gmail.com}
\altaffiltext{2}{Institute of Physics and Astronomy, University of
Potsdam, 14476 Potsdam, Germany}
\altaffiltext{3}{Physics
Department, Ben-Gurion University, Be'er-Sheva 84105, Israel;
eichler@bgu.ac.il} 
\begin{abstract}
We investigate the temporal and spectral correlations between flux and
anisotropy fluctuations of TeV-band cosmic rays in the 
light of recent data taken with IceCube. We find that for a conventional 
distribution of cosmic-ray sources the dipole anisotropy
is higher than observed, even if source discreteness is taken into account.
Moreover, even 
for a shallow distribution of galactic cosmic-ray sources and a reacceleration model,
fluctuations arising from source discreteness provide a probability only 
of the order of 
10\% that the cosmic-ray anisotropy limits 
of the recent IceCube analysis are met. 
This probability estimate is nearly independent of the exact 
choice of source rate, but generous for a large halo size. 
The location of the intensity maximum far from the Galactic Center is
naturally reproduced.
\end{abstract}
\keywords{cosmic rays}

\section{Introduction}
The anisotropy in the arrival directions of cosmic rays in the TeV
band has received renewed attention in recent years. A variety of experiments
\citep{2006Sci...314..439A,2007PhRvD..75f2003G,2009ApJ...692L.130A,
2009ApJ...698.2121A,2010cosp...38.2707Z,2010ApJ...718L.194A,2012ApJ...746...33A,
2012arXiv1202.3379D}
have reported results on the amplitude of the first harmonic in the siderial
anisotropy, which is typically found to be slightly less than $0.1\%$. All
experiments see only part of the sky, and it is not easy to reconstruct an
allsky dipole anisotropy, largely because the energy dependence of the
acceptance is different among the various detectors. In any case, the true
dipole anisotropy can be slightly higher than the sidereal first harmonic, i.e.
$\lesssim 0.1\%$. In addition, significant small-scale anisotropy was observed
that has not met an accepted explanation to date, although it is tempting to 
attribute it
to the local structure of the turbulent magnetic field in the
Galaxy \citep[e.g.][]{2011arXiv1111.2536G,2012PhRvL.108z1101G}.

Here we investigate whether or not the dipole anisotropy can be reproduced with
models of galactic cosmic-ray propagation that are tuned to fit available data
on secondary-to-primary ratios and the survival fraction of unstable isotopes.
For that purpose we use a Green-function method that solves a time-dependent
diffusion equation, similar to \citet{2012JCAP...01..011B}. Diffusion equations
of this type 
are derived by averaging the original Fokker-Planck transport equation for the
isotropic part of the cosmic-ray distribution function. By construction,
anisotropy in this treatment arises only from the diffusive (and possibly drift
or convective) flux and is always dipolar in nature. Modeling the small-scale
anisotropy, on the other hand, would be described with a 
transport equation derived from the original Fokker-Planck equation for the 
anisotropic part of the cosmic-ray distribution and is not considered here.

Our time-dependent treatment permits accounting for a discrete nature of
cosmic-ray source. A similar technique was used to model variations in the
spectrum of cosmic-ray electrons \citep{pe98,2003A&A...409..581P,
2011JCAP...02..031M}
and cosmic-ray ions \citep{2005ApJ...619..314B,2006AdSpR..37.1909P,
2012JCAP...01..010B,2012JCAP...01..011B}. We test various cosmic-ray 
source distributions in galactocentric radius, using the appropriate transport
parameters, and compute the expected temporal evolution of the anisotropy. This
approach permits a realistic assessment of the likelihood to find a certain
anisotropy for a specific choice of parameters. The 
occasionally discussed mean variance of the anisotropy is a considerably less
useful quantity, because the distribution of anisotropy amplitudes is highly
skewed.

We are particularly interested in the temporal and spectral
correlations between flux and anisotropy fluctuations. Our approach is ideally
suited to recover these correlations and investigate their effect, when we
compute flux and anisotropy over a finite spectral acceptance and compare with
observational results.

\section{Calculation of particle spectra}
\subsection{Cosmic-ray ions}

The differential density of TeV-band cosmic rays obeys a
continuity equation that can be written as
\begin{equation}
\frac{\partial N}{\partial t} - \frac{1}{r^2}\frac{\partial}{\partial r}
\left(r^2\,D\,\frac{\partial N}{\partial r}\right)=Q(E)\,\delta (t)\,
\frac{\delta (r)}{4\pi\,r^2}
\label{diff-eq}
\end{equation}
Here, $D=c\,\lambda_{\rm mfp}/3$ is the spatial diffusion coefficient, $r$ is
the distance from
the source, and $Q(t,E)$ is the differential production rate at the source.
Note that we neglect stochastic reacceleration, advection, and energy losses.
The first and the last of that list should only be important at GeV energies
and below, at least for cosmic-ray protons. The case of advection is less clear,
but probably also less important than it may be in the GeV band on account of
the energy dependence of diffusive propagation.

Using standard methods \citep{1962SvA.....6..317K,1964ocr..book.....G} one
derives
a general solution to the transport equation, assuming impulsive injection of
particle at time $t=0$,
\begin{equation}
N(r,t,E)= \frac{\Theta(t)}
{\left(4\pi\,D\,t\right)^{3/2}}\,Q(E)\,\exp\left(-\frac{r^2}{4\,D\,t}\right)
\label{n-eq}
\end{equation}
If more than one source contributes to the local cosmic-ray
flux at any time, their individual
contributions must be calculated using equation~\ref{n-eq} and then summed.

We now modify the solution to account for a finite size and lifetime
of sources and for escape from the Galaxy. It is known that assuming a point
source in space and time can lead to singularities in the cosmic-ray density
\citep[e.g.][]{2011JCAP...02..031M}. 
Real cosmic-ray sources, whatever their nature, have a finite spatial extent and
confine the accelerated particles for some time, for the acceleration is not
instantaneous. The most simple method to represent extended sources is to assume
that they have the shape of a thin, spherical shell of radius $R$. The
correspondence to the shell of a young supernova remnant is obvious, but the
applicability of the ansatz is not limited to SNR. In any case, averaging over
the shell yields
\begin{equation}
N(r,t,E)= \frac{\Theta(t)}{\left(4\pi\,D\,t\right)^{3/2}}\,Q(E)
\,\left[\frac{D\,t}{r\,R}\right]\,\sum_\pm\,(\pm 1)\
\exp\left(-\frac{(r\mp R)^2}{4\,D\,t}\right)\ .
\label{n-eq1}
\end{equation}
Note that this solution still has a singularity for $t\rightarrow 0$ at $r=R$,
which will disappear when we average over a finite lifetime, because
asymptotically 
$N\propto t^{-0.5}$ for $t\rightarrow 0$.

We do need to account for source size in computing the anisotropy, which is
given by the diffusive flux,
\begin{equation}
\delta=-\frac{\lambda_{\rm mfp}}{N}\,\frac{\partial N}{\partial r}
=\frac{\lambda_{\rm mfp}}{r}+\frac{3\,r}{2\,c\,t}-\frac{3\,R}{2\,c\,t}\,
\coth\left(\frac{r\,R}{2\,D\,t}\right)\ .
\label{eq-aniso}
\end{equation}
For small argument of the $\coth$ term, the first and last term on the RHS cancel each other.

For the integration over the time of source activity, we choose the most simple
form 
of a temporal profile, namely a constant source rate over an age interval
$[T_1;T_2]$ with the constraint that both $T_1$ and $T_2$ be positive.
The time integral can be easily performed analytically,
but leads to very small differences of exponentials and error functions that are
more difficult to numerically implement than a simple numerical integration.
A fact that we shall employ below is that either escape is irrelevant or the
effects of the finite extent of lifetime of the sources are negligible, provided
the source lifetime is much smaller than the escape time.

Diffusive escape from the Galaxy can be modeled using Dirichlet boundary
conditions at some distance from the midplane of the Galaxy, $N(z=\pm H)=0$.
Obviously, the boundary conditions break the spherical symmetry of the problem.
Whereas for very rare sources analytical approximations may ease the treatment 
\citep{2011ApJ...742..114P}, in the TeV band we expect cosmic-ray sources to
appear with a rate 
$\sim 1/(100\ {\rm yr})$, implying a relatively small distance to the nearest
source. Consequently, we attempt a full solution by using the method of mirror
sources. Since we have two boundaries, an infinite series of mirror sources with
alternating sign is 
needed \citep{2012JCAP...01..010B}. 

For the mirror sources only the Gaussian factor in equation
\ref{n-eq1} is relevant. It is accurate to second order in the ratio of
source lifetime and escape time to use an average age, $\tau = (T_2+T_1)/2$.
Mirror sources modify the z-dependent Gaussian factor in Equation~\ref{n-eq1} as
\begin{equation}
\exp\left[-\frac{(z-z_s)^2}{4\,D\,\tau}\right] \ \longrightarrow\ \\
G=\sum_{n=-\infty}^\infty \ 
(-1)^n\,\exp\left[-\frac{\left[z-2nH-(-1)^n z_s\right]^2} {4\,D\,\tau}\right]
\end{equation}
which, as is described in Appendix~\ref{appendix1},
can be approximated with good accuracy as
\begin{equation}
G\simeq \exp\left[-\frac{(z-z_s)^2}{4\,D\,\tau}\right]\,
\left(1+1.5\,x\right)^{1.25}\,\exp\left[-(1.5\,x)^{0.97}\right]
\qquad \qquad x=\frac{2\,D\,\tau}{H^2} .
\end{equation}

\subsection{Cosmic-ray electrons}

It is well known that the flux and spectrum of cosmic-ray electrons from
discrete sources are particularly sensitive to fluctuations
\citep{pe98,2003A&A...409..581P},
but the consequences for the electrons anisotropy are rarely discussed in the
literature.
In fact, the anisotropy in the intensity of cosmic-ray electrons at energies
between 60~GeV
and 480~GeV has only recently been measured \citep{2010PhRvD..82i2003A}. In the
case of electrons, we
need to account for their energy losses, but may neglect escape, at least above
$\sim$60~GeV.
Hence, we may describe the propagation of 
electrons at energies higher than a few GeV
by a simplified, time-dependent transport equation
\begin{equation}
{{\partial N_e}\over {\partial t}} - {{\partial}\over {\partial E}} (bE^2\, N_e)
-D\,E^a\, \nabla^2 N_e = Q_e
\label{transport}
\end{equation}
where we consider continous energy losses by synchrotron radiation and inverse 
Compton scattering, a diffusion coefficient $D_0\,E^a$ dependent on energy, and 
a source term, $Q_e$. Green's function for this problem can be
found in the literature \citep{1964ocr..book.....G}.
\begin{equation}
G_e={{\delta\left( t-t' +{{E-E'}\over {b\, E\, E'}} \right)}
\over {bE^2\, \left( 4\pi\,\lambda\right)^{3/2} }}\ 
\exp \left(-{{(r-r')^2}\over {4\, \lambda}}
\right)
\end{equation}
with
\begin{equation}
\lambda= {{D_0\,\left(E^{a-1} -E'^{a-1}\right)}\over {b\,(1-a)}}
\end{equation}

As in the case of cosmic-ray ions we assume the individual sources inject a
spectrum
$\propto E^{-s}$ at constant rate for a defined period of time, and so the
average over age from
$T_1$ to $T_2$ is
\begin{equation}
N_e=q_e\, E^{-s}\ \int_0^{\frac{1}{bE}}\ dt'\ 
{{\Theta (t'-T_1)\, \Theta (T_2 -t')}\over
{\left(4\pi \Lambda\right)^{3/2}\ ( 1-bEt')^{2-s}}}
\ \exp\left(-{{r^2}\over
{4\Lambda}} \right)
\end{equation}
where
\begin{equation}
\Lambda= {{D_0\,E^{a-1}}\over {b\,(1-a)}}\ \left(1-(1-bEt')^{1-a}\right)
\end{equation}
and $r$ is the distance between source and observer.

\begin{figure}
\includegraphics[width=0.5\columnwidth, keepaspectratio]{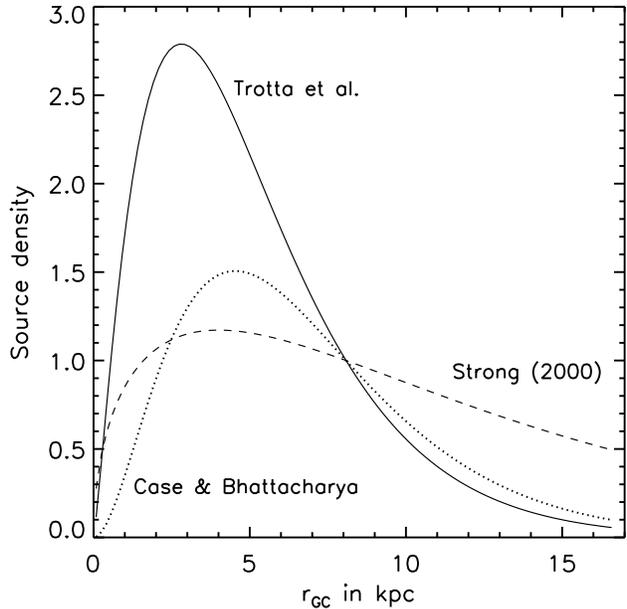}
\figcaption{Distribution of the cosmic-ray source density in galactocentric
radius, according to 3 estimates used in the literature. All curves are
normalized to unity at the solar circle. \label{pe:f1}}
\end{figure}

As in the case of ions, we can easily average over a shell of radius $R$ and
obtain
\begin{equation}
N_e=q_e\, E^{-s}\ \int_0^{\frac{1}{bE}}\ dt'\ 
{{\Theta (t'-T_1)\, \Theta (T_2 -t')}\over
{\left(4\pi \Lambda\right)^{3/2}\ (
1-bEt')^{2-s}}}\,\left[\frac{\Lambda}{r\,R}\right]
\,\sum_\pm\,(\pm 1)\ \exp\left(-{{(r\pm R)^2}\over {4\Lambda}} \right) ,
\label{eq-e-av}
\end{equation}
A general analytical solution is very difficult to obtain. However, an explicit
integration is 
unnecessary if $(T_2-T_1)\ll T_1$, provided the energy-dependent upper limit of
integration is observed.
If $s>2$, as is the case for cosmic-ray electrons, the integrand approaches zero
at 
$t\rightarrow (bE)^{-1}$. Then it is accurate to second order in $(T_2-T_1)/T_1$
to set
\begin{equation}
N_{av,e}=q_e\, E^{-s}\\ 
\frac{1}
{\left(4\pi \Lambda\right)^{3/2}\ (
1-bE\tau)^{2-s}}\,\left[\frac{\Lambda(\tau)}{r\,R}\right]
\,\sum_\pm\,(\pm 1)\ \exp\left(-{{(r\pm R)^2}\over {4\Lambda(\tau)}} \right)
\end{equation}
where $\tau=(T_2+T_1)/2$. The energy-loss time of electrons at 500~GeV in a
10-$\mu$G magnetic field 
is about $3\cdot 10^5$~years, much longer than the source life time $T_2-T_1$
that we consider here. If $bEt$ is a small parameter, we can expand the integrand in
Equation~\ref{eq-e-av}. Noting that for $bEt\ll 1$
\begin{equation}
\Lambda\simeq D_0\,E^a\,t\ ,
\end{equation}
we find that Equation~\ref{eq-e-av} assumes the same form as the time integral
over 
Equation~\ref{n-eq1}, the corresponding expression for cosmic-ray ions. 

\section{Intermittency and anisotropy}

We now run 100 simulations for each parameter combination.
Unless the number of cosmic-ray sources is very small, and hence intermittency
effects dominate, the source distribution is a critical factor in anisotropy
studies. Often, a function of the type
\begin{equation}
Q(r_{\rm GC})=\left(\frac{r_{\rm GC}}{R_\odot}\right)^a\,
\exp\left(-b\frac{r_{\rm GC}-R_\odot}{R_\odot}\right)
\end{equation}
is used to represent the density distribution in galactocentric radius, 
$r_{\rm GC}$. The parameters $a$ and $b$ derived by fitting observed
distributions of pulsars \citep{2004IAUS..218..105L}, 
supernova remnants \citep{1998ApJ...504..761C},
or other perceived tracers of cosmic-ray acceleration.

\begin{deluxetable}{llc}
\tablecolumns{3}
\tablewidth{0pc} 
\tablecaption{Standard parameters, mostly taken from the study of
\citet{2011ApJ...729..106T}.}
\tablehead{\colhead{Parameters}&\colhead{Symbol}&\colhead{Value}}
\startdata
Injection index& s&2.4\\
Energy dependence of diffusion  & $\delta$&0.3\\
Source distribution& a&1.25\\
Source distribution& b&3.56\\
Source rate& $P_Q$&$10^{-2}$ yr$^{-1}$\\
Source lifetime& &$2\cdot10^{3}$ yr\\
Source radius& $R$&$10$ pc\\
\enddata
\label{t1}
\end{deluxetable}

It has been noted earlier that the observed intensity of diffuse Galactic
gamma rays appears to require a substantially flatter source distribution
\citep{2000ApJ...537..763S}, although a metallicity gradient, and hence a
gradient in the $X_{CO}$ conversion factor of CO line intensity,
could explain the gamma-ray data as well \citep{2004A&A...422L..47S}. The 
comprehensive analysis of Fermi-LAT data by \citet{2012ApJ...750....3A}
also suggests a gradient in $X_{CO}$,
albeit a weaker trend as in \citet{2004A&A...422L..47S}. A similar conclusion was reached 
by \citet{2011ApJ...726...81A}.

Here we select from the literature three models of the cosmic-ray source
distribution in the Galaxy. Our sample comprises the relatively flat distribution used by
\citet{2000ApJ...537..763S}, the SNR distribution derived by
\citet{1998ApJ...504..761C}, which is slightly flatter than that of
\citet{2012arXiv1208.3107G}, and the pulsar distribution as presented by 
\citet{2011ApJ...729..106T}, which is close to that of 
\citet{2004IAUS..218..105L}. Figure~\ref{pe:f1} permits a visual comparison 
of the three models.

\begin{figure}
\includegraphics[width=0.5\columnwidth, keepaspectratio]{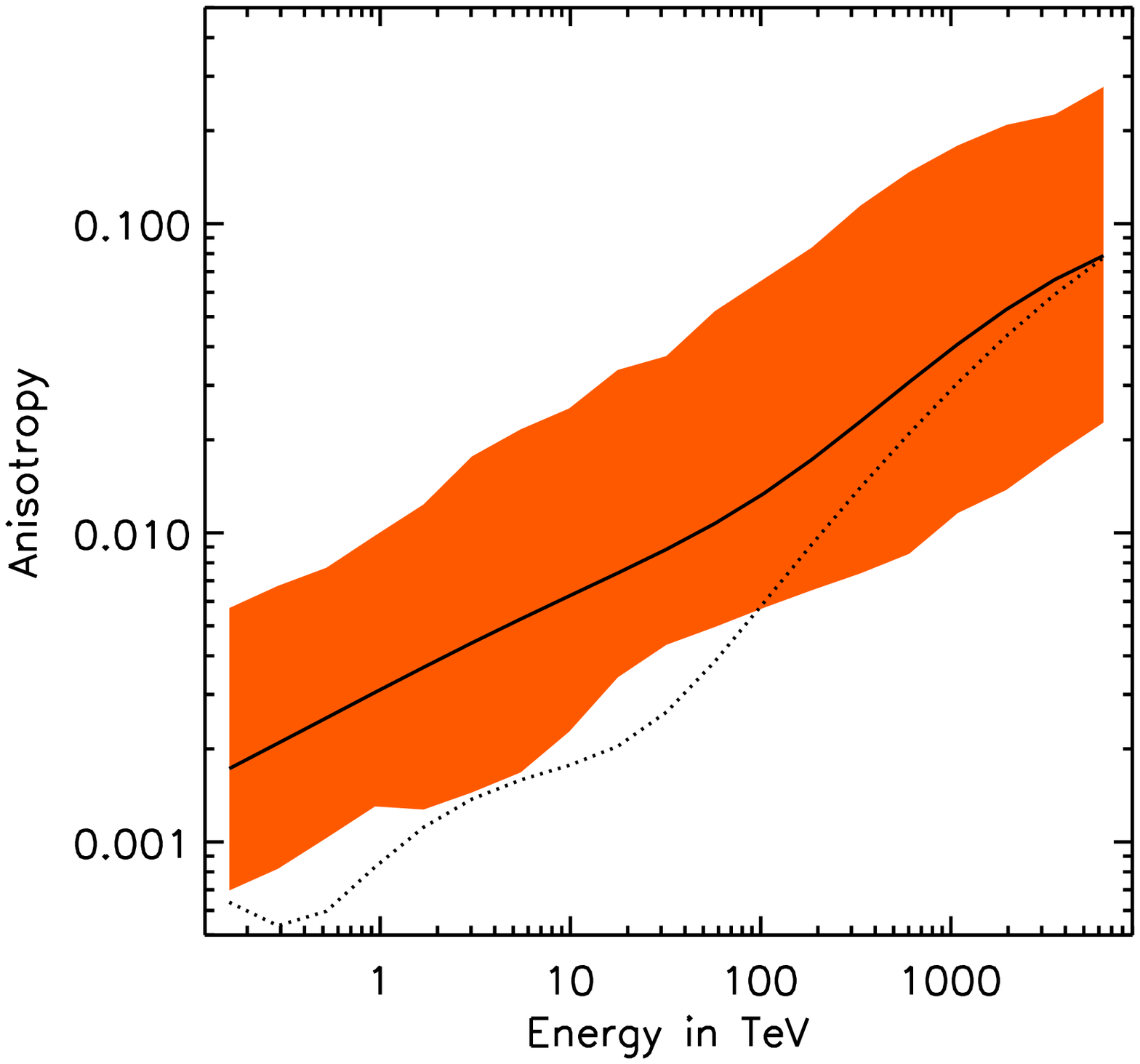}
\figcaption{Anisotropy for standard parameters (Trotta et al.)
and $H=5$~kpc. The red band
indicates the central 90\% of the fluctuation range. The solid line marks
the median of the distribution, the dotted line describes
the anisotropy amplitude in a randomly selected run.
\label{pe:f2}}
\end{figure}

According to \citet{2011ApJ...729..106T}, which we choose for the definition of
standard parameters, we have well-constrained propagation parameters,
such as the injection index
$s=2.4\pm 0.1$, and for the diffusion coefficient
\begin{equation}
D=D_0\,\left({E\over {\rm 4\ GeV}}\right)^\delta
\label{dif-tr}
\end{equation}
with $\delta =0.3\pm 0.05$ in reacceleration models, whereas a degeneracy arises
with the
halo size. In units of kpc, $3\le H\le 9$, and
\begin{equation}
D_0\simeq \left(1.2+1.3\,H\right)\cdot 10^{28}\ {\rm cm^2 s^{-1}}
\end{equation}
within about 20\% error margin. We try 4 values of H and derive the
95\%-confidence minimum of anisotropy in the 10-100 TeV band. Standard
parameters are summarized in Table~\ref{t1}. Note that these parameters were determined using the sharply peaked source distribution labeled \textit{Trotta et al.} in Figure~\ref{pe:f1}. The anisotropy and its variation is
shown as function of the particle energy in Figure~\ref{pe:f2}, full results are
detailed in Table~\ref{t2}. To be noted from Figure~\ref{pe:f2} is the skewness
in the distribution of anisotropy levels, which renders the use of a mean
fluctuation amplitude in the anisotropy less than useful.

\begin{deluxetable}{llrrrr}
\tablecolumns{6}
\tablewidth{0pc} 
\tablecaption{Proton anisotropy at 20 TeV total energy
for various parameters and 4 choices of halo size, $H$ in kpc.}
\tablehead{\colhead{Parameters}&\colhead{Confidence}&\colhead{H=3}&\colhead{H=5}
&\colhead{H=7}&\colhead{H=9}}
\startdata
Standard& 95\% lower limit&0.29\%&0.34\%&0.55\%& 0.46\%\\
Standard& Median& 0.60\%& 0.72\%& 0.81\%& 0.95\%\\
$\delta=0.25$&95\% lower limit&0.18\%&0.19\%&0.35\%& 0.35\%\\
$\delta=0.25$& Median& 0.40\%& 0.48\%& 0.54\%& 0.60\%\\
Case \& Bhattacharya&95\% lower limit&0.14\%&0.24\%&0.23\%& 0.29\%\\
Case \& Bhattacharya& Median& 0.44\%& 0.55\%& 0.62\%& 0.70\%\\
Strong (2000)&95\% lower limit&0.087\%&0.07\%&0.14\%& 0.12\%\\
Strong (2000)& Median& 0.38\%& 0.35\%& 0.37\%& 0.43\%\\
\enddata
\label{t2}
\end{deluxetable}

It is obvious that using the viable propagation parameters of
\citet{2011ApJ...729..106T}, the limits on anisotropy may be met only if either
the TeV-band cosmic rays are relatively heavy or the source distribution is very
flat. A shallow source distribution is certainly possible, but was not the one used
to find the propagation parameters best suited to reproduce B/C or $^{10}$Be
data. 

\begin{deluxetable}{llc}
\tablecolumns{3}
\tablewidth{0pc} 
\tablecaption{Parameters for a shallow source distribution, mostly taken
from Putze et al.}
\tablehead{\colhead{Parameters}&\colhead{Symbol}&\colhead{Value}}
\startdata
Injection index& s&2.415\\
Energy dependence of diffusion  & $\delta$&0.235\\
Source rate& $P_Q$&$10^{-2}$ yr$^{-1}$\\
Source lifetime& &$2\cdot10^{3}$ yr\\
Source radius& $R$&$10$ pc\\
\enddata
\label{t1f}
\end{deluxetable}

\begin{figure}
\includegraphics[width=0.5\columnwidth, keepaspectratio]{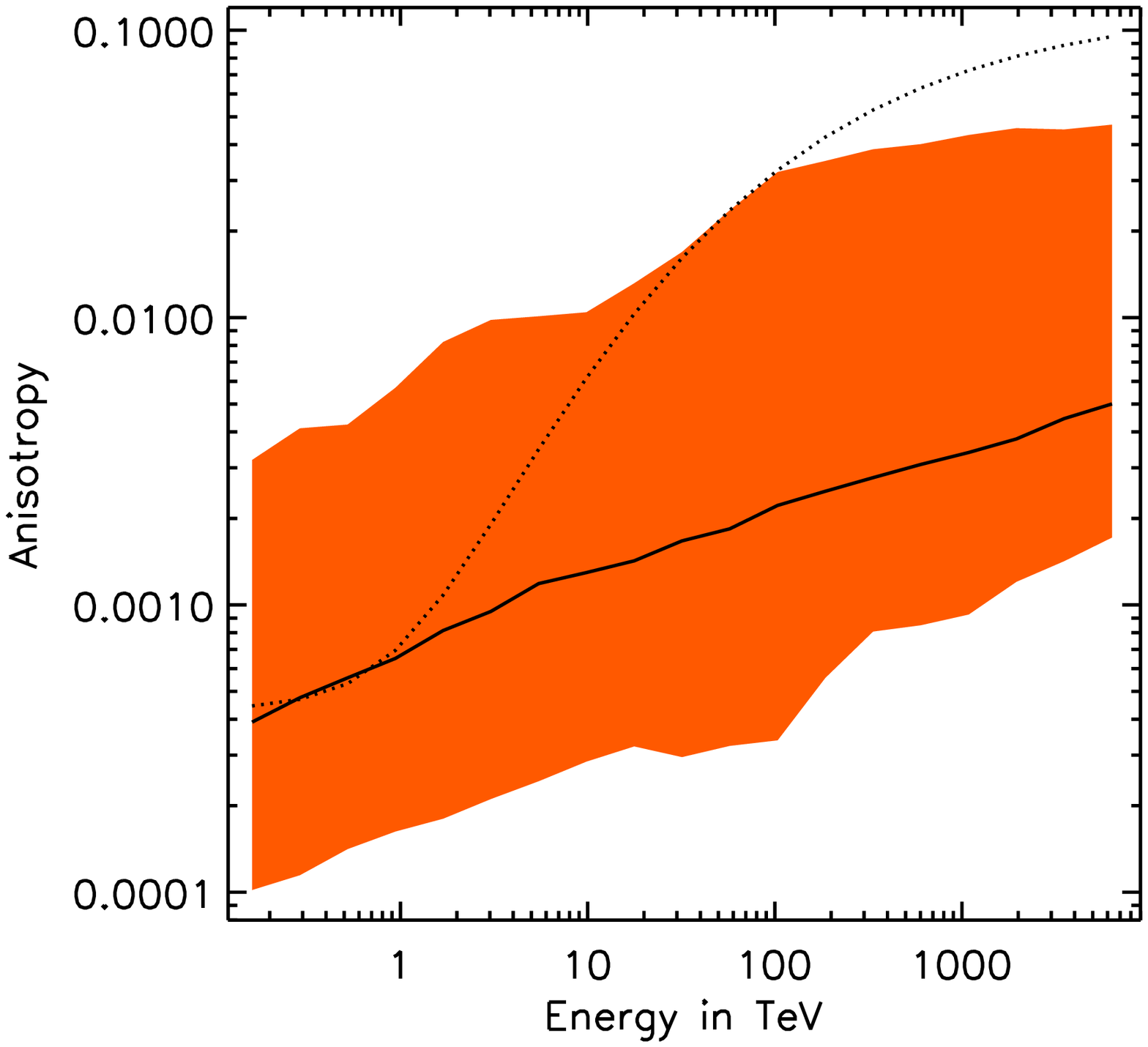}
\figcaption{Anisotropy for a shallow source distribution with parameters
as listed in Table~\ref{t1f}
and halo size $H=4.5$~kpc. As in Figure~\ref{pe:f2}, 
the red band indicates the central 90\% of the
fluctuation range, the solid line marks
the median of the distribution, the dotted line describes
the anisotropy amplitude in a randomly selected run.
\label{pe:f3}}
\end{figure}

\subsection{Shallow source distribution}
To verify the reliability of our findings, we test with results of 
\citet{2010A&A...516A..66P} who used a different propagation model and a
constant source density. We do not use their reacceleration/convection model,
because in 2-D geometry with
radial boundary the best-fit parameters indicate that advection has little
importance above a TeV.

For the reacceleration model the scaling of the diffusion coefficient with halo
size (in kpc) is 
\begin{equation}
D_0\simeq H\,1.04\cdot 10^{28}\ {\rm cm^2 s^{-1}}
\label{dif-fl}
\end{equation}
with $1.5\le H\le 6$. Numerically, the scale factor of the diffusion coefficient
is similar to that of \citet{2011ApJ...729..106T}, but at 20 TeV the diffusion
coefficient is actually a factor 2.6 smaller on account of the weaker energy
dependence of diffusion. Other parameters are summarized in Table~\ref{t1f}.
We use the relatively flat source distribution of \citet{2000ApJ...537..763S} out
to $r_{\rm GC}=20$~kpc, and henceforth refer to it as the \emph{shallow source 
distribution}.

\begin{deluxetable}{llrrrr}
\tablecolumns{6}
\tablewidth{0pc} 
\tablecaption{Proton anisotropy at 20 TeV total energy
for a shallow source distribution.}
\tablehead{\colhead{Parameters}&\colhead{Confidence}&\colhead{H=1.5}&\colhead{
H=3} &\colhead{H=4.5}&\colhead{H=6}}
\startdata
Standard& 95\% lower limit&0.026\%&0.023\%&0.032\%& 0.043\%\\
Standard& Median& 0.12\%& 0.10\%& 0.14\%& 0.14\%\\
Source rate&95\% lower limit&0.022\%&0.037\%&0.03\%& 0.035\%\\
$P_Q=10^{-3}$ yr$^{-1}$& Median& 0.12\%& 0.13\%& 0.12\%& 0.16\%\\
Source rate&95\% lower limit&0.034\%&0.054\%&0.035\%& 0.054\%\\
$P_Q=10^{-4}$ yr$^{-1}$& Median& 0.13\%& 0.15\%& 0.13\%& 0.16\%\\
\enddata
\label{t2f}
\end{deluxetable}

Results are summarized in Table~\ref{t2f}. To be noted from the table is the
strong reduction in anisotropy resulting from the moderate (factor 2.6)
decrease in the diffusion coefficient (cf. Table~\ref{t2}, in particular the runs
for Strong (2000)). 

\begin{figure}
\includegraphics[width=0.4\columnwidth, keepaspectratio]{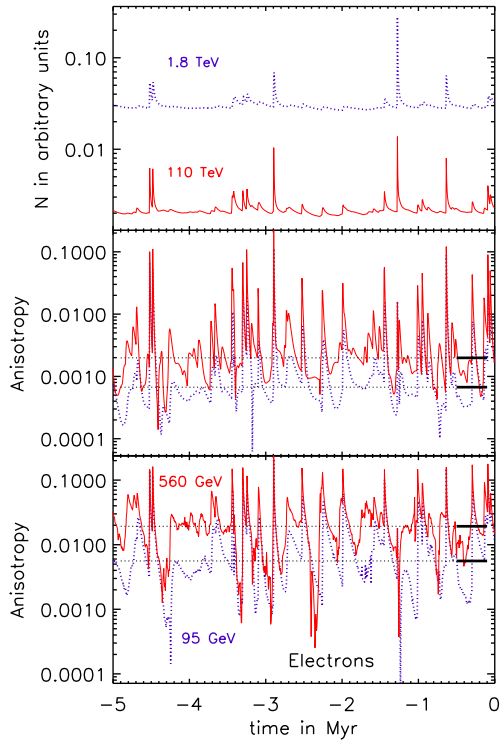}
\figcaption{Light curves of proton flux at 2 energies (top panel),
proton anisotropy (middle panel), and electron anisotropy at two lower energies
(bottom panel). The black bars and dotted horizontal lines indicate the median
of the anisotropies.
\label{pe:f4}}
\end{figure}

\section{Light curves}
\subsection{General behavior}
Having established that a rather flat source distribution is necessary to
explain the low anisotropy amplitude measured at multi-TeV energies, we now
discuss light curves. We want to infer the temporal and spectral
correlations between flux and anisotropy fluctuations, which must be taken into
account when 
computing the flux and anisotropy over a finite spectral acceptance and comparing with
observational results.

A number of insights can be gained by first discussing light curves at specific
energies. Figure~\ref{pe:f4} shows the temporal variations of proton flux,
proton anisotropy, and electron anisotropy for the standard parameters with halo
size $H=3$~kpc and a shallow source distribution (see Table~\ref{t1f}). Our
findings can be summarized as follows:

\begin{figure}
\includegraphics[width=0.5\columnwidth, keepaspectratio]{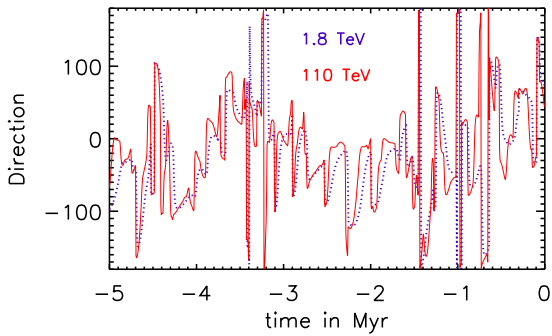}
\figcaption{Distribution in Galactic longitude of the
direction of maximum proton intensity.
\label{pe:f5}}
\end{figure}

\begin{itemize}
\item The proton flux displays small-amplitude fluctuations with rare,
short-lived spikes, a result already discussed by \citet{2005ApJ...619..314B}.
\item Spikes in flux are generally accompagnied by spikes in anisotropy at all
energies and reflect the appearance with nearby, young cosmic-ray sources. The events 
may change the diffusion coefficient on account of streaming instabilities, in which 
case this calculation would overestimate the anisotropy during spikes.
\item Deep, short-lived dips can be observed in the anisotropy, which have no
counterpart in the flux and do not appear at all energies. These dips arise
from a random balancing of the contributions of the dominant cosmic-ray sources
and are therefore a hallmark of discrete sources. 
\item There is little correlation between the small-amplitude fluctuations in
the anisotropy at different energies, which implies that care must be exercized
when comparing with airshower data, which carry contributions from cosmic-rays
over a wide range of energies.
\item The predicted electron anisotropy is lower than the upper limits published
by \citet{2010PhRvD..82i2003A}, but higher by a factor $\sim$5 than the GALPROP
predictions shown in their Figure 9. GALPROP does not properly account for
discrete sources and therefore would underpredict the anisotropy.
\end{itemize}

The direction of maximum intensity is almost always in the Galactic plane, 
because the distribution of cosmic-ray sources with the vertical coordinate,
$z$, is very narrow. In Figure \ref{pe:f5} we show how the anisotropy direction 
varies with time, using the same simulation as for Figure~\ref{pe:f4}
for ease of comparison. To be noted from the figure is the correspondence of the
deep dip in anisotropy at -3.2~Myr with a 360-degree swing of the anisotropy
direction. Generally, the variations in the anisotropy direction are large and
in general not achromatic, although the difference in directions at 1.8~TeV and
110~TeV rarely exceeds 30 degrees.  

Ground-based observatories do not see the entire sky and therefore resort to
measuring the first harmonic in right ascension, while integrating over the 
declination band that is accessible to them. The projection of a 
first harmonic in galactic longitude depends on the direction of maximum
intensity and on the declination band in question.

\begin{figure}
\includegraphics[width=0.5\columnwidth, keepaspectratio]{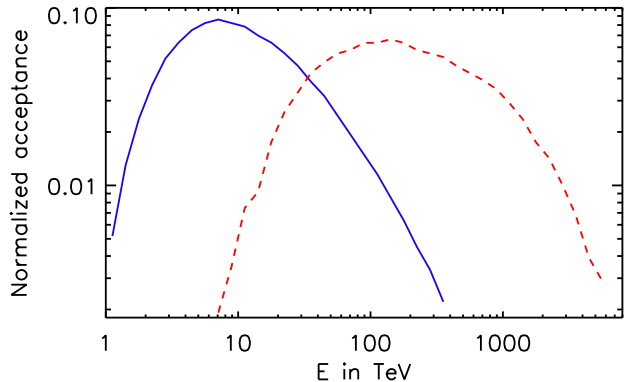}
\figcaption{The distribution in energy of the primary particle
for the high-energy (red line) and low-energy (blue line) event selections,
derived under the assumption that all primary cosmic rays are protons.
\label{pe:f6}}
\end{figure}

\begin{figure}
\includegraphics[width=0.5\columnwidth, keepaspectratio]{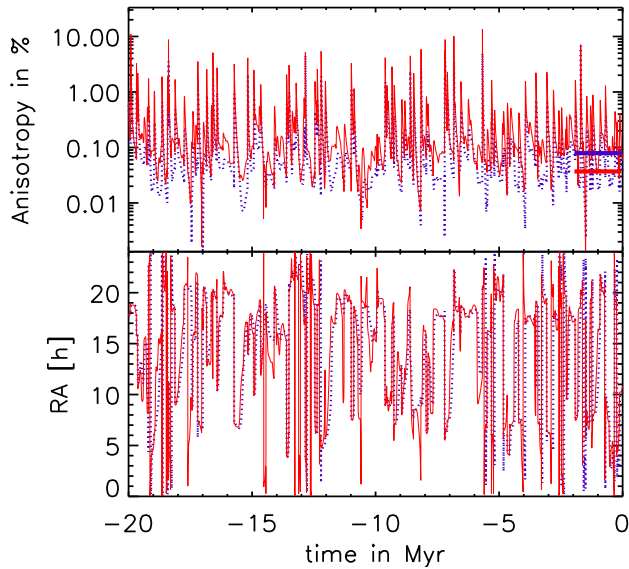}
\figcaption{Top panel: Expected sidereal anisotropy in IceCube data at declination
$\theta_c=-45^\circ$ for the 
high-energy selection (in red) and low-energy band (in blue). 
The bars on the right indicate measured limits. The source rate is assumed to
be 
one per thousand years. Bottom panel: Location in Right Ascension of the
maximum intensity in IceCube data for the 
high-energy selection (in red) and low-energy band (in blue). 
\label{pe:f7}}
\end{figure}

\subsection{Comparison with IceCube results}
The IceCube collaboration has presented cosmic-ray anisotropy measurements in
two
broad energy bands \citep{2012ApJ...746...33A}, based on a total of about 
30 billion cosmic-ray-induced muon events. Airshowers caused by heavy primaries 
approximately behave like the superposition of airshowers produced by the 
individual nucleons. Therefore, the detection with IceCube of a muon with a
certain energy requires a heavy primary of much higher energy than it would, if
the primary were a proton. In terms of propagation, heavy cosmic rays with 
charge number $Z$ and energy $E$ will behave like protons of energy $E/Z$.
Thus, 
light and heavy cosmic-ray nuclei that have similar propagation properties, and 
hence similar anisotropy, will also account for muons of similar energy in 
IceCube. We have seen
in Figure~\ref{pe:f3} that a factor 2 in energy on average yields a 20-\%
change 
in anisotropy. We have also seen that the spectrum of cosmic rays shows little 
variation, except possibly during intense spikes in flux.
It is therefore a good approximation to only consider cosmic-ray 
protons. 

The IceCube collaboration was kind enough to provide us with the distribution
in energy of the primary particle
for their 2 muon event selections, derived under the
assumption that all primary cosmic rays are protons. These distributions are 
also shown in
Figure~\ref{pe:f6}. Obviously the median primary
particle
energies are lower than for the mixed composition assumed in 
\citet{2012ApJ...746...33A}, but these are the distributions we need to combine
with our proton-propagation calculation described in the preceding sections. 

IceCube reports the sidereal first harmonic in the cosmic-ray intensity
averaged over the declination range $-25^\circ < \theta < -72^\circ$. Obviously,
the dependence of acceptance on the primary energy is not constant over this
declination range, but we will ignore that and project the cosmic-ray dipole 
anisotropy onto the sidereal band at the characteristic declination 
$\theta_c=-45^\circ$, which is approximately the median declination of events 
accepted in the IceCube analysis. Note that we compute the direction of
the cosmic-ray dipole anisotropy (cf. Figure~\ref{pe:f5}), which displays
considerable variability, and so does its projection onto the sidereal band.

A dipole anisotropy of amplitude $\delta$ pointing at declination $\theta_d$
provides a sideral first harmonic at declination $\theta_{\rm c}$ of amplitude
\begin{equation}
\delta_{\rm proj} (\theta_{\rm c})=
\frac{\delta\,\cos\theta_d\,\cos\theta_{\rm c}}
{1+\delta\,\sin\theta_d\,\sin\theta_{\rm c}}
\end{equation}
The denominator can essentially always be set to unity. We simulate the
cosmic-ray intensity for $50$~Myr to obtain a statistically meaningful data set.
In Figure~\ref{pe:f7} we show the expected anisotropy in the IceCube data over 
$20$~Myr only for a source rate $P_Q=3\cdot 10^{-3}\ {\rm yr^{-1}}$.

The IceCube collaboration reports the following sidereal anisotropies:
\begin{eqnarray}
\delta_{\rm obs}({\rm low\ energy})&=
(7.9\pm 0.1_{\rm stat}\pm 0.3_{\rm sys})\cdot 10^{-4} \nonumber\\
\delta_{\rm obs}({\rm high\ energy})&=
(3.7\pm 0.7_{\rm stat}\pm 0.7_{\rm sys})\cdot 10^{-4}
\end{eqnarray}
The expected anisotropy is simultaneously below both values for 8.2\% of the 
simulated time period of $50$~Myr. 
Even if we account for the uncertainties and set the anisotropies to 
$8.2\cdot 10^{-4}$ and $4.8\cdot 10^{-4}$, respectively, implying that
individually there is only a 16\% probability that the true anisotropy
is higher, they are met at only 12.9\% of the time.

Also shown in Figure~\ref{pe:f7} is the location of peak intensity in the sidereal band studied by IceCube. To be noted from the figure is that in particular during anisotropy lows the peak intensity is found far from the Galactic-Center direction at
RA$\simeq 17.8$h and not necessarily coincident for the 2 energy bands. Both results 
reproduce the IceCube observations. 

\begin{deluxetable}{llrrrr}
\tablecolumns{6}
\tablewidth{0pc} 
\tablecaption{Summary of anisotropy results. The first two columns give the assumed 
source rate and halo size. The next three columns give the probability, that the 
anisotropy is below that measured with IceCube, and the anisotropy median at low energy 
and high energy. The last column identifies the corresponding lines in 
Figure~\ref{pe:f8}. }
\tablehead{\colhead{Rate}&\colhead{Halo}&\colhead{Prob.}&\colhead{Med. LE}&\colhead{
Med. HE} &\colhead{Line} \\
\colhead{$\left[10^3\,{\rm Myr^{-1}}\right]$}&\colhead{$\left[{\rm 
kpc}\right]$}&\colhead{$\left[\%\right]$}&
\colhead{$\left[10^{-4}\right]$}&\colhead{$\left[10^{-4}\right]$}&\colhead{}}
\startdata
\hline
0.3& 3&8.2&7.1&13.5& Black \\
0.3& 7&3.3&8.3&15.7& Blue dotted \\
3& 7&4.2&7.7&15.2& Red dotted \\
3& 3&8.2&6.8&12.8& Red \\
1& 3&10.4&6.8&13.2& Blue \\
10& 3&9.1&6.2&12.0& Green \\
\enddata
\label{t5}
\end{deluxetable}

\begin{figure}
\includegraphics[width=0.5\columnwidth, keepaspectratio]{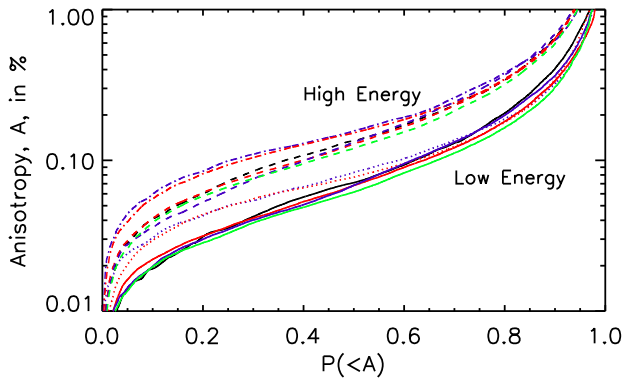}
\figcaption{The integral probability, $P(<A)$, to measure an anisotropy lower than
a certain threshold, $A$. Solid and dotted lines refer to the low-energy data selection
of IceCube, dashed and dot-dashed line are for the high-energy data. 
The parameters of each curve are listed in table~\ref{t5}. 
\label{pe:f8}}
\end{figure}

To verify that this results does not significantly depend on the assumed source rate
and halo size, we have varied the rate between $P_Q=3\cdot 10^{-4}\ {\rm yr^{-1}}$
and $P_Q=3\cdot 10^{-2}\ {\rm yr^{-1}}$, and the halo size between 
$H=3$~kpc and $H=7$~kpc. In no case was the probability to simultaneously meet 
both anisotropy values higher than 11\%. Table~\ref{t5} summarizes the results, and 
Figure~\ref{pe:f8} displays the integral probability to measure an anisotropy lower than
a certain threshold, $A$. To be noted from the figure is that varying the source rate 
does not change the probability distribution in a significant manner, but choosing a 
large halo size appears to reduce the probability of a low anisotropy. 

As last test, we 
repeated one simulation ($H=3$~kpc and $P_Q = 3\cdot 10^{-3}\ {\rm yr^{-1}}$) for a perfectly
flat source distribution out to $r_{\rm GC}=20$~kpc. The probability of meeting the IceCube
upper limits was with 11\% only marginally er than for the shallow source distribution of 
\citet{2000ApJ...537..763S}.

\section{Discussion}

This is not the first calculation of cosmic-ray anisotropy in the TeV band.
The main advantage of this study lies in the proper accounting for the correlations
between flux and anisotropy variations, a restriction to cosmic-ray propagation parameters
that were deemed viable in dedicated studies, and an accurate folding with the acceptance
distribution of a recent observational study.

We have specifically compared with recent measurements with the IceCube experiments
\citep{2012ApJ...746...33A}, which leads us to the following conclusions:

\begin{itemize}
\item Consistent with earlier findings,
a cosmic-ray source distribution in the Galaxy that fits to the deduced populations
of pulsars or SNR, combined with propagation parameters that reproduce
the observed secondary-to-primary ratios, will lead to a dipole anisotropy that
is higher than observed, even if source discreteness is taken into account.
\item Also in agreement with earlier studies, a shallow energy dependence of diffusion
is necessary, thus requiring a reacceleration model to fit the GeV-band 
Boron-to-Carbon ratio. 
\item For a shallow source distribution, fluctuations arising from source discreteness provide a probability of the order of 10\% that the cosmic-ray anisotropy limits in the high-energy and low-energy bands
of the recent IceCube analysis can be simultaneously met, approximately independent of the exact choice of source rate, but less likely so for a large halo size.
\item The location of the intensity maximum far from the Galactic Center and at different locations for the 2 energy bands are naturally reproduced when source discreteness is taken into account.
\end{itemize}

It is unclear what type of cosmic-ray source in the Galaxy would have a 
distribution in the Galaxy as shallow as required in our study. The question 
arises whether or not modifications to the description of diffusive transport 
can render a steep source distribution viable.
The critical constraint appears to be imposed by very low dipole anisotropy
in the high-energy band of IceCube data, i.e. at approximately 100~TeV energy per 
nucleon. Is it possible that the extrapolation to this energy of the cosmic-ray 
propagation parameters does not hold?

Recent data suggest that cosmic-ray spectra
\citep{2010ApJ...714L..89A,2011Sci...332...69A}, and the Boron-to-Carbon ratio
\citep{2012ApJ...752...69O}, flatten above a few hundred GeV/nuc. 
At this time we do not have a 
prevalent interpretation of both the cosmic-ray
spectral hardening and the flat B/C ratio above a few hundred GeV/nuc.
The spectral
hardening of cosmic-ray primaries could simply be a fluctuation arising from source 
discreteness, but in that case the Boron-to-Carbon ratio should fall off more steeply, 
because the flux of cosmic-ray secondaries fluctuates very little 
\citep{2005ApJ...619..314B}. Spectral hardening can also reflect variations
in the production spectra of different sources
\citep{1972ApJ...174..253B,2001A&A...377.1056B,2011PhRvD..84d3002Y},
in which case the Boron-to-Carbon ratio would be unaffected.
If the upturn in the Boron-to-Carbon ratio is, on the 
other hand, due to the fragmentation of primaries and the re-acceleration of 
secondaries inside cosmic-ray sources \citep{2009PhRvL.103h1104M,2012A&A...544A..16T},
then the energy dependence of the diffusion 
coefficient cannot be argued to be very shallow. The only possibility 
that would imply a very low anisotropy is a change of the energy
dependence of the diffusion coefficient 
at a few hundred GeV/nuc, e.g. from $\delta=0.25$ to $\delta=0.15$ 
\citep{2012ApJ...752...68V,2012A&A...544A..16T}.
Speculative though that may be, the extrapolation
using $\delta=0.15$ to $10^{17}$~eV may render viable the notion that a significant
fraction of the sub-ankle cosmic-ray flux is galactic in origin 
\citep[e.g.][]{2011ApJ...742..114P}.
  
Recently, \citet{2012PhRvL.108u1102E} suggested a common solution to the anisotropy
and cosmic-ray gradient problems, that involves a strong toroidal magnetic field
in the Galaxy.
Diffusion perpendicular to the azimuthal magnetic field (with coefficient
$D_\perp\propto E^{0.6}$, as opposed to $D_\parallel\propto E^{0.3}$) would
be the dominant transport mechanism,
assuming field-line wandering and large-scale velocity turbulence can be ignored. 
The authors further posit
that $D_\perp (r_{\rm GC})\propto 1/D_\parallel(r_{\rm GC})$ and 
$D_\perp\propto Q(r_{\rm GC})^\tau$, where $Q(r_{\rm GC})$ is the density of
cosmic-ray sources and $\tau \simeq 0.85$. At least up to 1 TeV particle energy, the 
predicted radial anisotropy appears to be commensurate with measured values.

We posit that the scenario would do little, if anything, to resolve the 3-D anisotropy
problem for discrete sources (the authors only discuss anisotropy for continuous 
sources in a 2-D scenario). In the direction of fast diffusion,i.e. along the 
large-scale toroidal magnetic field, we see particles of the same age from sources at 
larger distance, thus increasing their contribution to the anisotropy. The volume 
sampled, and thus the number of sources contributing to the local cosmic-ray flux, 
will also increase, thus reducing the likelihood of a fluctuation toward low 
anisotropy, but not efficiently enough to overcome anisotropy arising from 
source discreteness. In the scenario of \citet{2012PhRvL.108u1102E} we
therefore expect significantly more anisotropy in
direction of the ordered magnetic field than in radial direction.

\acknowledgements
We thank Rasha Abbasi and the IceCube collaboration for providing the
proton-only acceptance distribution for their anisotropy analysis.
MP acknowledges support by the Helmholtz Alliance for Astroparticle Physics, HAP,
funded by the Initiative and Networking Fund of the Helmholtz Association.
DE acknowledges support from the Israel-U.S. Binational Science Foundation, the Israeli Science Foundation, and the Joan and Robert Arnow Chair of Theoretical Astrophysics.

\appendix
\section{Analytical appoximation of escape}\label{appendix1}

Mirror sources modify the z-dependent term as
\begin{equation}
\exp\left[-{{(z-z_s)^2}\over {4\,D\,\tau}}\right] \ \longrightarrow\ \\
G=\sum_{n=-\infty}^\infty \ 
(-1)^n\,\exp\left[-\frac{\left[z-2nH-(-1)^n z_s\right]^2} {4\,D\,\tau}\right]
\end{equation}
that can be rewritten as
\begin{equation*}
G=\exp\left[-{{(z-z_s)^2}\over {4\,D\,\tau}}\right]\ \times 
\end{equation*}
\begin{equation}
\times\left(1+\sum_{n\neq 0} \ (-1)^n\,\exp\left[{{2\,n\,H\,\left[z-(-1)^n
z_s-n\,H\right]
-z\,z_s\,(1-(-1)^n)}\over 
{2\,D\,\tau}}\right]\right)
\end{equation}
The solution is inconvenient, because one needs to sum over many terms to find
convergence for all $t$. We are interested in the intermittency behaviour and
thus need to account for many million cosmic-ray sources in the Galaxy, which
makes computational efficiency imperative.
We can expand the exponential in the series to give
\begin{equation*}
f=\exp\left[{{4nH\,\left[z-(-1)^n
z_s\right]-4n^2H^2-2\,z\,z_s\,(1-(-1)^n)}\over 
{4\,D\,\tau}}\right]
\end{equation*}
\begin{equation}
\simeq \exp\left[-{{n\,H\,\left[(-1)^n z_s + n\,H\right]}\over 
{D\,\tau}}\right]\ \times\,\left(1+z\,A + {{z^2\,A^2}\over 2} +\dots\right)
\end{equation}
where
\begin{equation}
A={{2\,n\,H -z_s\,(1-(-1)^n)}\over {2\,D\,\tau}}
\end{equation}
The last factor in brackets is essentially unity, unless $H^2\gg D\,\tau$ in
which case
the exponential yields zero, because its argument is always at least a factor
$H/z$ larger 
as the z-dependent terms in the last factor.

We may now expand the remaining exponential for small $z_s$ to find
\begin{equation}
f\simeq \exp\left[-{{n^2\,H^2}\over 
{D\,\tau}}\right] \times\left(1- (-1)^n \,z_s\,{{n\,H}\over {D\,\tau}} + {1\over
2}\,
\left[{{z_s\,n\,H}\over {D\,\tau}}\right]^2 +\dots\right)
\end{equation}
where summation over $\pm n$ removes the first-order term and otherwise yields a
factor 2.
Hence
\begin{equation}
G\simeq \exp\left[-{{(z-z_s)^2}\over {4\,D\,\tau}}\right] \times 
\left[1+2\,\sum_{n=1}^\infty \ (-1)^n\, \exp\left[-{{n^2\,H^2}\over 
{D\,\tau}}\right]\ \times\,\left(1 + {1\over 2}\,
\left[{{z_s\,n\,H}\over {D\,\tau}}\right]^2 \right)\right]
\end{equation}
As for $z$, the second-order term in $z_s$ can become relevant only when the
exponential 
is practically zero and the term in brackets unity. Therefore, it can be
neglected.
A nice analytical approximation for the term in brackets is given by
\begin{equation}
G\simeq \exp\left[-{{(z-z_s)^2}\over {4\,D\,\tau}}\right]\,
\left(1+1.5\,x\right)^{1.25}\,\exp\left[-(1.5\,x)^{0.97}\right]
\qquad \qquad x={{2\,D\,\tau}\over {H^2}}
\end{equation}
which deviates from the true solution by less than 7.2\% for $x\le 8$, for which
the term
in brackets is down to $3.6\cdot 10^{-4}$.

\end{document}